\documentclass[english]{emulateapj}
\usepackage[latin9]{inputenc}
\usepackage{graphicx}
\usepackage{amssymb} 
\usepackage{esint}



\usepackage{babel}

\begin{document}

\title{The Triple evolution dynamical instability:\\
 Stellar collisions in the field and the formation of exotic binaries }

\author{Hagai B. Perets$^{1,2}$ \& Kaitlin M. Kratter$^2$}

\affil{$^1$ Deloro Fellow, Technion - Israel Institute of Technology, Haifa, Israel 32000 \\$^2$ Harvard-Smithsonian Center for Astrophysics, 60 Garden St.; Cambridge,
MA, USA 02138}
\begin{abstract}
Physical collisions and close approaches between stars play an important
role in the formation of exotic stellar systems. Standard theories
suggest that collisions are rare, occurring only via random encounters
between stars in dense clusters. We present a different formation
pathway, the triple evolution dynamical instability (TEDI), in which
mass loss in an evolving triple star system causes orbital instability.
The subsequent chaotic orbital evolution of the stars triggers close
encounters, collisions, exchanges between the stellar components,
and the dynamical formation of eccentric compact binaries (including 
Sirius like binaries).
We demonstrate that the rate of stellar collisions due to the TEDI
is approximately $10^{-4}$ yr$^{-1}$ per Milky-Way Galaxy, which
is nearly $30$ times higher than the total collision rate due to random
encounters in the Galactic globular clusters. Moreover, we find that the dominant
type of stellar collisions is qualitatively different; most collisions
involve asymptotic giant branch stars, rather than main sequence,
or slightly evolved stars, which dominate collisions in globular clusters.
The TEDI mechanism should lead us to revise our understanding of collisions
and the formation of compact, eccentric binaries in the field. 
\end{abstract}

\section{Introduction}

The predominant cause of stellar collisions is thought
to be random encounters between stars in dense clusters. These collisions,
where stars pass within a single stellar radius of each other, may
result in a wide range of outcomes, most notably the formation of
close and eccentric compact binaries \citep[and references therein]{sha99,sha02,dav02,iva10}.
Because the collision rate per star, $\Gamma$, due to random encounters
depends directly on the local number density of stars $\Gamma\propto n$,
such interactions are common only in the densest stellar environments
such as globular clusters (GCs). Outside of these clusters, the densities
are lower by a factor as great as $10^{7}$: random collisions effectively do not occur in
the field. In this paper, we describe a dynamical instability that
occurs in evolved systems containing three or more stars%
\footnote{We focus here on triple systems, but the scenario is also applicable
to higher multiplicity systems.%
}. This mechanism, which we term the triple evolution dynamical instability
(hereinafter TEDI), 
leads to close encounters and collisions between stars even in low
density stellar environments.

The TEDI occurs due to an interplay between stellar mass loss and
orbital instability. Consider a hierarchical triple system composed
of an inner binary coupled to an outer binary, where the outer binary
consists of a tertiary companion in orbit about the center of mass
of the inner pair. When the orbits of the inner and outer pair are
well separated, the system can remain stable indefinitely. Stellar
evolution can disrupt this architecture. When the more massive component
of the inner binary evolves, it begins to shed mass. As a result,
the orbits expand in proportion to the ratio between the initial and
final mass in the enclosed system \citep{had+63,egg06}. Since the
relative mass loss in the inner binary is greater than that in the
outer binary, the inner orbit encroaches on the outer orbit. This
relative orbital expansion can trigger a dynamical instability, which
results in the dissolution of the system (see Fig. \ref{fig:scenarios}).

The subsequent evolution of such unstable systems is very similar
to that of binary-single star encounters in globular clusters, and
often leads to close encounters between any of the three components.
The TEDI provides a unique pathway to physical collisions during these
close approaches: as stars evolve off of the main sequence, they not
only lose mass, but also expand radially by a factor of a few hundred
on the Asymptotic Giant Branch (AGB). The simultaneity of mass loss
(and thus instability) with stellar bloating greatly increases the
cross section for collisions. For this reason, as we now demonstrate,
the TEDI is the dominant route to stellar collisions in the Universe,
producing nearly 30 times more collisions than random gravitational
encounters in GCs.

The significance of mass-loss induced instability in triple systems
has been proposed in several studies \citep{kis+94,ibe+99,per10,fre+11,por+11}.
However, the frequency of this phenomena, and its importance for stellar
collisions and the consequences thereof, have not been explored. 

Our primary goal in this paper is to estimate the rate of collisions per galaxy  
induced by the TEDI mechanism, $\Gamma_{\rm col}$.  This calculation requires the estimation of the following
quantities:
\begin{itemize}
\item \textbf{$N_{\star}$} - The number of stars in the Galaxy
\item \textbf{$f_{{\rm evolve}}$} - The observed fraction of evolved stars
\item \textbf{$f_{{\rm triple}}$} - The fraction of stars in hierarchical
triple systems
\item \textbf{$f_{{\rm des}}$} - The fraction of hierarchical systems that
become unstable as the stars evolve
\item \textbf{$f_{\rm col}$} - The fraction of unstable triple systems that
end with a collision between two of the stars
\end{itemize}
Thus $\Gamma_{\rm col}=N_{\star}\times f_{\rm evolve}\times f_{\rm triple}\times f_{\rm des}\times f_{\rm col}$.

Observations provide estimates of the first two variables, and less certain estimates for the third (triple fraction). Attaining
estimates for $f_{{\rm des}}$ and $f_{\rm col}$ occupies the remainder of this paper, as well as addressing the uncertainties in the triple fraction, $f_{{\rm triple}}$. The latter is done with both observational estimates and theoretical population synthesis models to provide independent estimates. 
The main difficulties in doing so are due to the limited data existing on
triple systems and the distribution of their properties, and the large
phase space of possible evolutionary routes. Given these caveats,
we  can  only hope to provide an order of magnitude estimate of the 
importance of the TEDI.  Though we provide the formal statistical uncertainties
for the resulting fractions we obtain,  the
systematic uncertainties due to our limited knowledge of the distribution
of triple system properties are likely larger than the statistical ones.

The outline of the paper  is as follows. In Section \ref{sec:TEDI} we
quantify the criterion for the TEDI. We then determine the fraction
of systems that undergo the TEDI, $f_{\rm des}$, in Section \ref{sec:basic-estimate},
followed by an estimate of the frequency, $f_{\rm col}$, of collisions in the destabilized
systems (\S\ref{sec:coll}). In Section \ref{sec:rate}
we use these results to provide an estimate of the total collision
rate in the galaxy. We describe the various consequences of the TEDI
in Section \ref{sec:outcomes}, and briefly point out the influence of secular
dynamics on triple stellar evolution (\S\ref{sec:koz}). We close
the paper with the discussion and summary in Section \ref{sec:summary}.

\section{The triple evolution dynamical instability (TEDI) }

\label{sec:TEDI} A triple system is dynamically unstable when $Q<Q_{{\rm st}}$,
where $Q=a_{{\rm out}}(1-e_{{\rm out}})/a_{{\rm in}}$ and $Q_{{\rm st}}$
is (we adopt \citeauthor{val+08} 2008; but see also \citeauthor{mar08} 2008 for a different definition) : \begin{equation}
Q_{{\rm st}}=3(1+m_{3}/M_{12})^{1/3}(1-e_{{\rm out}})^{-1/6}\left(\frac{7}{4}+\frac{1}{2}\cos{i}-\cos^{2}i\right)^{1/3},\label{eq:stability}\end{equation}
 where $a_{{\rm in}}$ and $a_{{\rm out}}$ are the semi-major axes
(SMAs) of the inner and outer binaries in the triple system, $e_{{\rm out}}$
is the outer binary eccentricity, and $m_{3}$ and $M_{12}=m_{1}+m_{2}$,
are the masses of the outer component and the inner binary, respectively.
The mutual inclination between the inner and outer binaries is $i$.

As the primary (with mass $m_{1}$) in the inner binary evolves off
the main sequence (MS), it loses mass. We consider only systems where
the stellar-wind driven mass loss occurs adiabatically, i.e. on a
timescale that is long compared to the outer orbital period (prompt
mass loss could occur in supernovae and also lead to instability)
\footnote{The triple supernova instability scenario is limited only
to high mass systems. Moreover, in the majority of cases, supernova
explosions lead to the prompt unbinding of the system (especially
if they are accompanied by a natal kick), making longer term chaotic
evolution, and hence physical collisions, less likely than in the
case of adiabatic mass loss. The contribution of this process to the
over all collision rate in the Galaxy is therefore likely to be small,
and not affect our main results. Nevertheless, this process, and the
effects of natal kicks, could be especially important for collisions
with neutron stars and black holes, and merits further study.}. We further assume that mass loss is isotropic, and that
mass transfer is negligible. The expansion of the inner orbit due
to mass loss will be larger than the expansion of the outer orbit
because the relative mass loss in the inner system is larger. Thus the SMAs evolve as: 

\begin{equation}\frac{a_{{\rm in,f}}}{a_{{\rm in,i}}}=\frac{m_{{\rm 1,i}}+m_{2}}{m_{{\rm 1,f}}+m_{2}}>\frac{m_{{\rm 1,i}}+m_{2}+m_{3}}{m_{{\rm 1,f}}+m_{2}+m_{3}}=\frac{a_{{\rm out,f}}}{a_{{\rm out,i}}},\label{eq:expansion}\end{equation}
 \citep{had+63} where the subscripts correspond to the system components
and the evolutionary stage of the system ($i$ and $f$ for the initial
and the final, post-mass loss system, respectively). The expansion
of each of the SMAs in direct proportion to the mass loss derives
from the conservation of the quantity $GMa(1-e^{2})$, where $M$
is the total system mass. Note that slow, isotropic mass loss is required
for $e$ to remain constant.

The ratio between the SMAs of the outer and inner binaries therefore
increases, by a factor of \begin{equation}\frac{{a_{{\rm in,f}}}/{a_{{\rm out,f}}}}{{a_{{\rm in,i}}}/{a_{{\rm out,i}}}}=\left(\frac{m_{{\rm 1,i}}+m_{2}}{m_{{\rm 1,f}}+m_{2}}\right)\left(\frac{m_{{\rm 1,f}}+m_{2}+m_{3}}{m_{{\rm 1,i}}+m_{2}+m_{3}}\right).\label{eq:sma-ratio-increase}\end{equation}
 The TEDI occurs when this increase drives a stable triple system
across the threshold defined in Eq. (\ref{eq:stability}). The subsequent
dynamical evolution of an unstable system is chaotic. The system may
evolve through various configurations, including exchanges between
the stars in the inner and outer binaries, eccentricity excitation,
and collisions (see Fig. \ref{fig:scenarios} for an example). The
final outcome is typically the ejection of one star and/or a physical
collision. Because this three body problem is genuinely chaotic (typical
Lyapunov exponent of $1/2$; \citealp{iva+91,hei+99,val+06} and references
therein), the outcome is strongly dependent on small variations in
the initial conditions.

\includegraphics[scale=0.5]{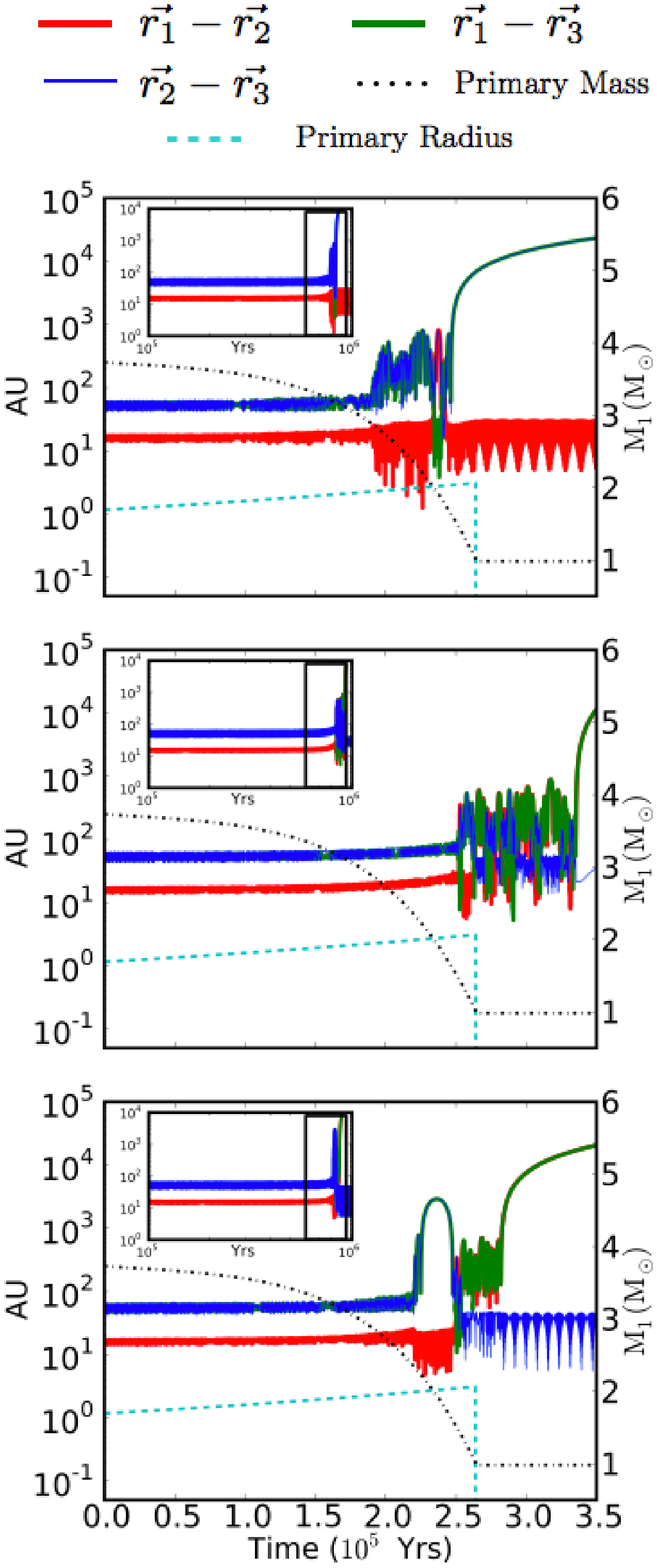}
\begin{figure}[h!]
\begin{centering}

\par\end{centering}

\centering{}\caption{\label{fig:scenarios}{Three-body simulations of the dynamical evolution
of triple systems undergoing adiabatic mass loss.} The separations
($r$) between each of the three components in each system are plotted
as a function of time. Also shown are the mass (dashed-dotted) and
the stellar radius (dashed) of the primary, $M_{1}$. Thick lines
show the pairs including the primary. Top (1a): A collision between
a MS star and an AGB star (the subsequent evolution, assuming point
masses and no collision is also shown). Middle (1b): the formation
of an eccentric close WD-MS binary, similar to the Sirius system.
Bottom (1c): the WD primary is ejected after several exchanges, leaving
behind an eccentric MS-MS binary. In all cases the systems started with masses of $M_1 = 5\, M_\odot,\, M_2 =2.0\, M_\odot,\, M_3 = 2.25\, M_\odot,\, a_{\rm in} = 15\,{\rm AU}, a_{\rm out} = 53\, \rm{ AU},\, e_{\rm out }=0.1,\, i= 0.1$.}

\end{figure}

\section{Estimate of the fraction of systems destabilized due to the TEDI}

\label{sec:basic-estimate} To determine whether a given system is
unstable, we simultaneously follow its orbital and stellar evolution,
and check whether or not it violates the stability criterion at any point during its evolution.
We estimate the fraction of destabilized systems in two ways:
(1) we determine the fraction of observed triple systems \citep{tok08}
with known orbital parameters that become unstable, and (2) we calculate the unstable fraction of
 a synthetic population of triples modeled on binary
statistics to account for potential biases in the observed sample.

We first describe the procedure to identify systems that should become
unstable, and then calculate the instability rates for the observed
and the synthetic samples.

\subsection{Identifying Unstable Systems}

\label{sec:unstable}

In order to determine whether a given triple system is susceptible
to the TEDI during its evolution, we make use of single and binary
population synthesis stellar evolution codes (SSE and BSE, described
in detail in \citealp{hur+00,hur+02}), to which we couple an analytic
calculation of the orbital evolution of the wider third component
using Equation (\ref{eq:expansion}).

The specific procedure is as follows. 
\begin{enumerate}
\item The inner binary in the triple is evolved using the BSE code up to
a Hubble time (13.7 Gyrs), assuming Solar metallicity. 
\item The third companion is evolved using the SSE code up to a Hubble time
(13.7 Gyrs), assuming Solar metallicity. 
\item The mass-dependent orbital evolution of the inner binary and outer
single star are combined to form a time series evolution of all three
stars. The inner orbit of the binary is determined by the BSE code.
The outer orbit is updated at every output time step, $t_i$ of the BSE using
Eq. (2), where the initial and final time steps are the preceding
and current time steps \[
\frac{a_{{\rm out}}(t_i)}{a_{{\rm out}}(t_{i-1})}=\frac{m_{{\rm triple}}(t_{i-1})}{m_{{\rm triple}}(t_i)}.\]
 We require that mass transfer between the inner binary and the third
companion is always negligible, so that we may apply the adiabatic
formula above. If the outer binary separation is smaller than $15$
AU, we assume non-negligible mass transfer is possible either
following a common-envelope phase for the inner-binary, or stellar
evolution of the tertiary. We therefore consider all triples with
outer separations smaller than $15$ AU to be stable to the TEDI. 
\item We calculate the stability coefficient, $Q$ at every step in the
evolution, and using Eq. (1) find all systems that become unstable.
Note that if the inner binary merges, or any of the stars explode
as supernovae prior to destabilization, the system is considered stable
(in the context of adiabatic mass loss). All other destabilized triples
comprise our sample of unstable triples whose
size is denoted by $N_{{\rm des}}$. 
\item For all destabilized systems we find the radius, $R_{{\rm des}}$,
of the mass losing evolving star at the point of destabilization,
and note the types and radii of the other components. 
\item The fraction of destabilized systems is defined as $f_{{\rm des}}={N_{{\rm des}}/N_{{\rm triple}}}$. 
\end{enumerate}

\subsection{Observed Triples}

The observed triple sample \citep{tok08} provides estimated masses
and orbital periods for each star, but does not provide their mutual
inclinations and eccentricities.  We randomly assign each system relative inclinations 
and eccentricities drawn from realistic distributions (see \citealt{fab+07}).

We employ a distribution of
eccentricities that is function of period, in line with
observed binaries \citep{duq+91}. For periods shorter than 1000 days,
the eccentricity is chosen from a Rayleigh distribution {[}$dp\propto e\exp(-\beta e^{2})de${]}
with $\left\langle e^{2}\right\rangle ^{1/2}=\beta^{-1/2}=0.33$.
For periods longer than 1000 days, the eccentricity is chosen from
an Ambartsumian distribution ($dp=2ede$), which corresponds to a uniform
distribution on the energy surface in phase space. The inclinations
are chosen from a random distribution that is uniform in $\cos i$.
We repeat the random sampling of these properties 30 times for each
system. 
 
We only evolve triples that are  initially stable, since unstable systems are unlikely to be
observable due to the instability timescale. If our random assignment of $e$ and $i$ produce
an unstable system, we draw new values from the distributions.

We follow the procedure discussed in Section \ref{sec:unstable} to determine
whether a given system becomes unstable. We find that the fraction
of all observed triple systems with primary mass $m_{1}>1$ M$_{\odot}$
destabilized within a Hubble time is $f_{{\rm des,o}}=0.035\pm0.006$,
where the error bars correspond to the 1 $\sigma$ statistical uncertainty.

\begin{figure}
\begin{centering}
\includegraphics[scale=0.35]{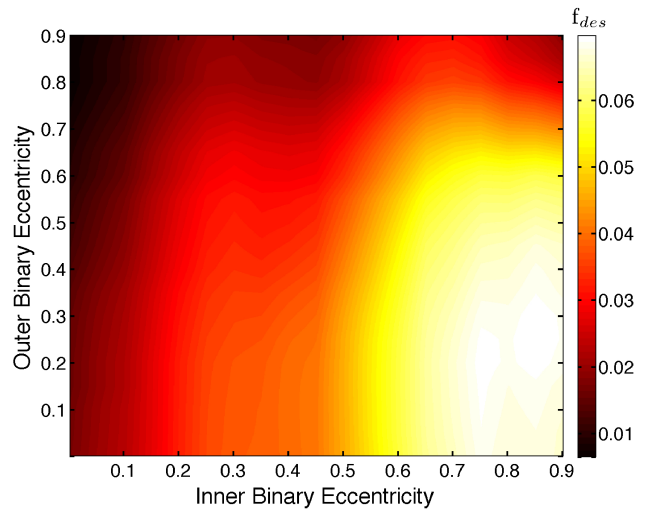}
\par\end{centering}

\centering{}\caption{\label{fig:unstable-eccs} The fraction of destabilized systems, $f_{\rm des}$,
is shown for a grid of inner and outer eccentricities applied to the
observed sample of triple systems (smoothed and interpolated; original
bin size is $\Delta e=0.05$).}

\end{figure}

In Fig. \ref{fig:unstable-eccs}, we illustrate the
sensitivity of the results to the eccentricity distribution: we calculate
the fraction of unstable systems for a complete grid of eccentricities.
For this sample we conservatively assume retrograde orbits, $i=180^{\circ}$,
which are the most stable configurations (see Eq. \ref{eq:stability}).

\subsection{Synthetic Sample of Triples}

We complement our use
of the existing observed triple sample, which may introduce unknown
biases, with a synthetic samples of triples. The method and
the assumptions used to produce the synthetic samples are based on
a method used by \citep{fab+07} for low mass triples and \citep{per09}
for high mass triples (with primary mass larger than $3$ M$_{\odot}$).

We assume that the orbital and stellar characteristics of triple systems
are equivalent to two (uncorrelated) binary systems, with the inner
binary acting like a single mass when choosing the third component.
The study by \citet{kra+11} suggests that this assumption is in good
agreement with observations (though this was only checked for a small
sample of triples). We choose initial orbital periods such that both
the inner and outer binaries follow the observed distribution. The
inner and outer periods are chosen from a distribution that is Gaussian
for low mass binaries with $M<3M_{\odot}$ \citep{rag+10} and log
flat for massive binaries corresponding to $f(r)\propto1/r$ i.e.,
$\ddot{O}$pik's law \citep{kob+07}.%
\footnote{We assume A stars share a similar distribution as that of F/G/K stars
in the \citep{rag+10} sample.%
}

The mass ratio of the inner binary, $q=m_{1}/m_{2}$, is chosen from
a Gaussian distribution (mean of $0.6$ and dispersion of $0.1$)
for low mass primaries and a power law for massive stars ($f(q)\propto q^{-0.4}$);.
The mass of the tertiary companion, $m_{3}$, is chosen such that
$q=m_{3}/(m_{1}+m_{2})$ follows the same mass ratio distribution.
Thus the mass of the third star is correlated with the combined mass
of the inner binary.

The inner and outer SMAs are computed from the masses and periods
assuming non-interacting Keplerian orbits. The inclination and eccentricity
distributions are chosen in the same manner as the observed sample.
We produce a sample of triples with primary masses in the range $1\le m_{p}\le20$,
with 1000 systems per primary mass bin (with bin size $\Delta M=0.3$
M$_{\odot}$, for a total number of $6\times10^{4}$ systems).
 Note that since the triple outer component is randomly chosen
from a distribution with respect to the inner binary mass, the term
primary mass refers to the inner binary primary mass (which is not
necessarily the most massive component in the triple).

\begin{figure}
\begin{centering}
\includegraphics[scale=0.35]{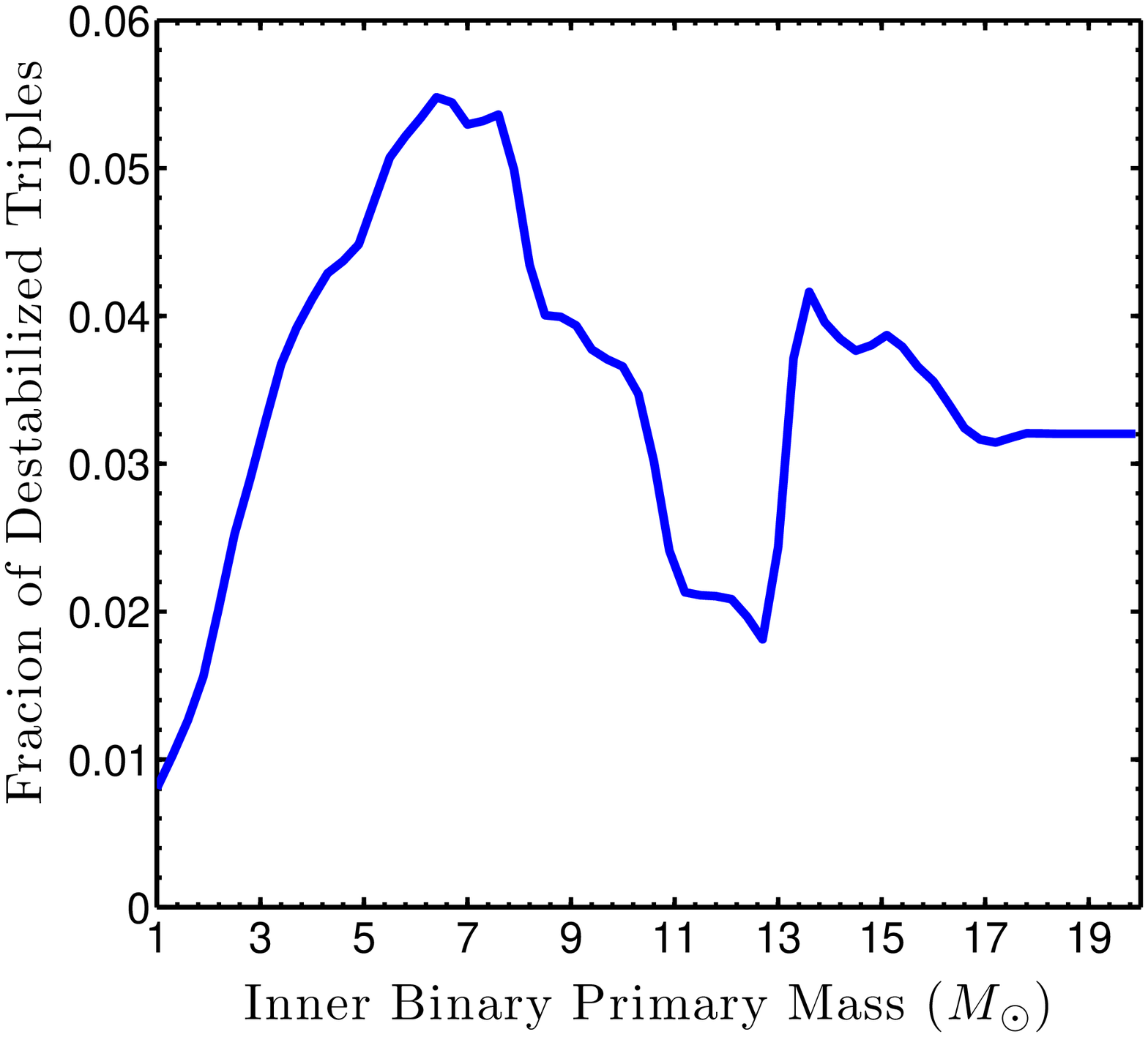}
\par\end{centering}

\begin{centering}
\includegraphics[scale=0.35]{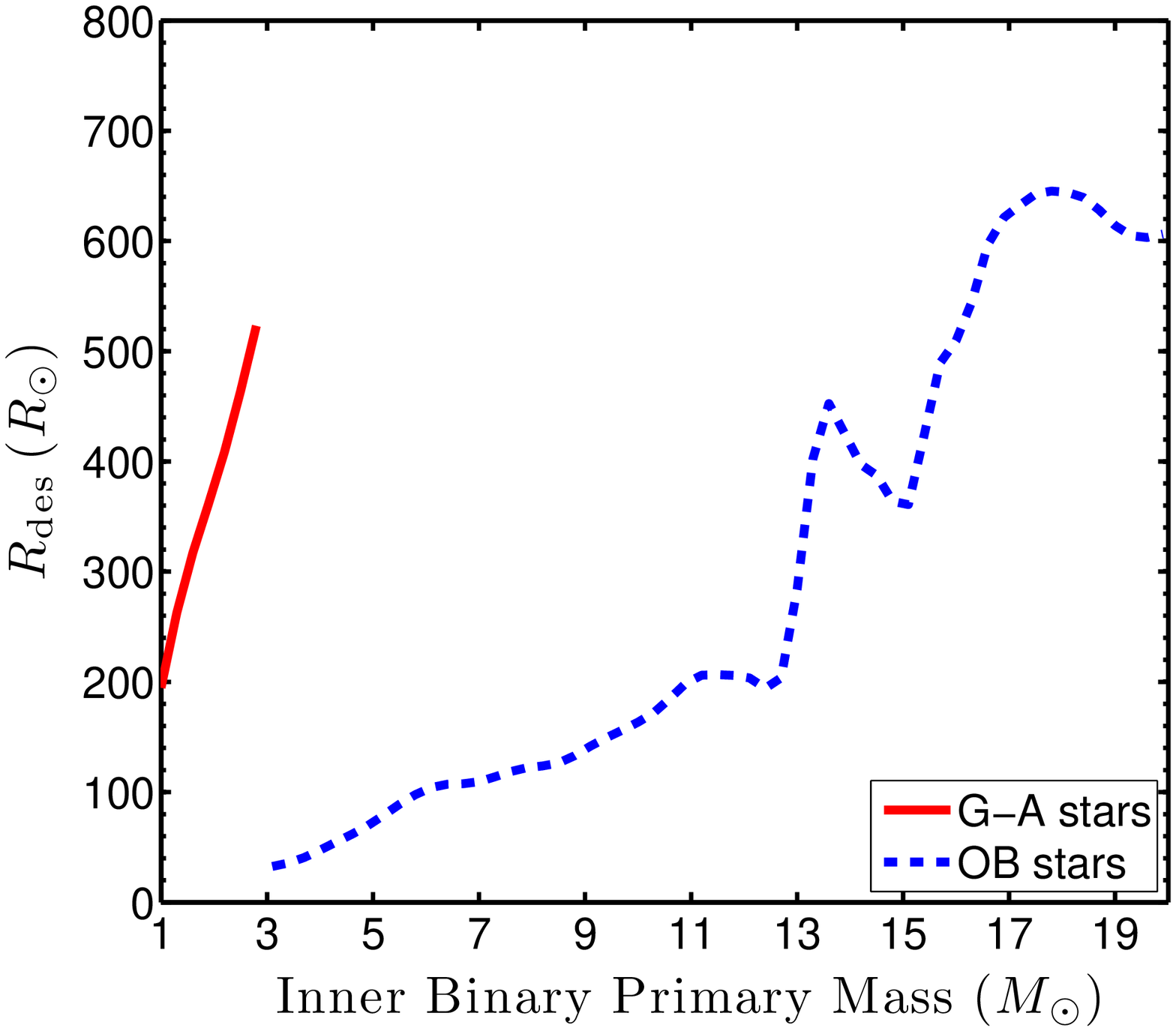}
\par\end{centering}

\centering{}\caption{\label{fig:TEDI-triples} Top: The fraction of destabilized triples
as a function of the inner binary primary component mass. The down
turn at about $8$ M$_{\odot}$ is due to supernova explosions that
occur before the system can be destabilized due to mass loss. Bottom:
The median radius of the mass losing stars at the onset of instability,
as a function of mass. Separate lines indicate the low mass ($m<3$
M$_{\odot}$) and the high mass stars $(m>3$ M$_{\odot}$), for which
different triple distributions are assumed (see text). }

\end{figure}

In Fig. \ref{fig:TEDI-triples}a we show the fraction of destabilized
triples from this sample as a function of primary mass. Fig. \ref{fig:TEDI-triples}b
shows the typical radius of the mass losing star at the time of destabilization
($R_{{\rm des}}$). The change in the assumed binary properties between
the low mass and the high mass stars introduces a sharp change in
the typical $R_{{\rm des}}$, since higher mass stars have more compact
triple configurations on average, and can be destabilized at an earlier
phase of their evolution, at which point the evolved star has a smaller
radius.

Using the fraction of destabilized triples as a function of primary
mass, $f_{{\rm des,s}}(m)$, shown in Fig \ref{fig:TEDI-triples},
we can estimate the total fraction of destabilized triples. The statistical
uncertainty for these fractions is small (1-2 \%; not shown). The
fraction of observed stars in triple systems 
ranges from 11\% for Solar like (F, G, K) stars \citep{rag+10,ras+10}
to possibly 50\% for more massive B stars \citep{eva11}. We use a
Salpeter initial mass function, 
$g(m)=dn/dm\propto m^{-2.35}$, and assume the triple fraction of
massive stars is about $5$ times larger than that of lower mass stars
\citep{kob+07,eva11}. Given these distributions, we estimate the
total fraction of destabilized triple systems in our synthetic sample
to be: 
\begin{eqnarray}
  &f_{{\rm des,s}}=\int_{1}^{3}f_{{\rm des,s}}(m)\frac{dn}{dm}dm\nonumber \\
  &+5\times\int_{3}^{30}f_{{\rm des,s}}(m)\frac{dn}{dm}dm=0.010+0.043=0.053.
\end{eqnarray}

The overall fraction of destabilized triples in our synthetic
sample is higher than but consistent with the results for the observed sample (0.053, compared
with 0.035) (see also Fig. \ref{fig:unstable-eccs}).
Our estimates suggest that a few percent of all evolved triple systems undergo the
TEDI. 

We now turn to direct integrations of three-body systems to both validate the analytic instability criterion and to determine the fraction
of unstable systems that produce a collision, $f_{\rm col}$.

\subsection{Three-body Simulations}

\label{sec:N-body}

Because TEDI systems begin in stable configurations and adiabatically
evolve into unstable configurations, they may not be directly comparable
to unstable triples explored in previous studies \citep{val+06}.
Therefore we conduct a series of three-body simulations including
mass loss both to check that our destabilized systems are truly unstable,
and to estimate the frequency of collisions based on the statistics
of close approaches during the chaotic evolutionary phase. In the appendix, we provide a
semi-analytic estimate of the collision rate.

 The integrations are done using a modified version
of the Hermite integrator described in \citet{hut+95}. The integrator
uses a variable time step constrained to be $10^{-2}$ of the minimum
collision time of any two stars. Reducing the time-step by a factor
of 10 does not change the outcome. In runs without mass loss, energy
is conserved to about one part in $10^{10}$ over the 10 Myr run time.
We only consider point masses, and include no tidal effects, or mass
transfer. We use a constant mass loss rate, determined by the total
mass loss in the system as found in the stellar evolution calculation,
divided by a constant mass loss timescale, $\tau_{{\rm loss}}=0.5$
Myr (the typical lifetime of the highly evolved stars, during which
most of the mass loss occurs). After the mass loss epoch, $\tau_{{\rm loss}}$,
we continue the simulation, at constant mass up to 10 Myrs to check
for longer term stability. In cases where our stellar evolution calculation
shows that instability occurred only after both the primary and secondary
in the inner binary evolved, we simulated the system beginning only
at the second stage of mass loss (i.e. from the secondary), after
the primary became a white dwarf.

We simulated 100 realizations of each observed triple system found
to be unstable in Section \ref{sec:basic-estimate}; each realization
differed only in the initial orbital phase of the stars. Since we
do not account for mass transfer, we simulated only systems in which
the inner binary separation was larger than 15 AU (these comprise
nearly half of the observed sample). We ran $300$ different system
configurations, resulting in a total of $3\times10^{4}$ runs.

To identify systems that become unstable, we require that either one
star is ejected from the system, or an exchange occurs between the
tertiary and one of the inner binary components. We find that
approximately $0.33$ of the systems found to be unstable in our simplified
calculations become unstable in our simulations over a 0.5 Myr run.
Although more triples are likely to show exchanges and escapes over
longer time scales (for example we find approximately $55$\% were destabilized
within 10 Myrs), we adopt the short timescale simulation result, since our primary concern is with collisions, which are
much  more likely if they occur while the primary is on the AGB. 

We take the most conservative estimate possible by assuming that the results from the three body integrations
are more representative of the entire sample. We thus reduce the fraction of unstable observed 
systems found in Section \ref{sec:basic-estimate} ($f_{\rm des,o}=0.035$) by a factor of $0.33$; we
adopt as the fraction of destabilized systems $f_{{\rm des}}=0.035\times0.33\simeq0.012$.

\section{Stellar collisions during the TEDI (calculating $f_{\rm col}$)}

\label{sec:coll} Having calculated the expected fraction of triple
systems that are unstable to the TEDI, we now estimate the fraction
of such systems that might suffer physical collisions. We show in the appendix that applying  the results of previous simulations for both single star - binary encounters and unstable triple systems
 gives the fraction of unstable systems that undergo a collision as $0.49\pm0.03$.  The analytically derived collision rate for unstable triples is significantly higher than that for random encounters because an unstable triple typically undergoes multiple close approaches before the system dissolves, and because the stellar radius at the time of collision is so large. Because the analytic results are based on initially unstable systems covering  a limited part of the the three body problem phase-space, we determine a more conservative collision rate based on our three body integrations (see below).

\subsection{Collisions in three-body simulations}

\begin{figure}
\begin{centering}
\includegraphics[scale=0.45]{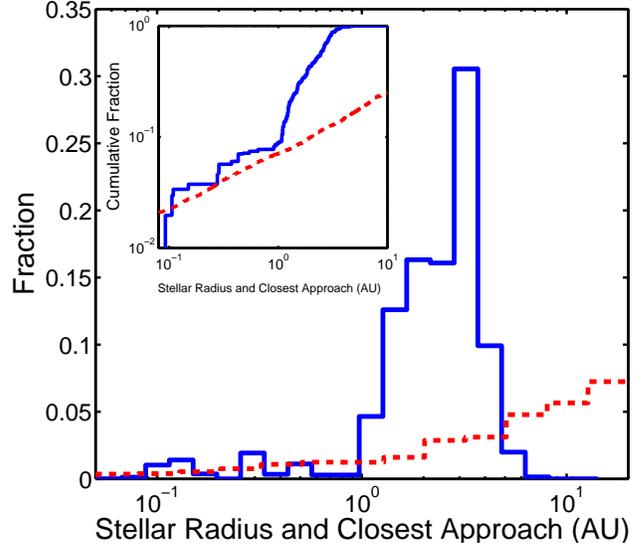}
\par\end{centering}

\centering{}\caption{\label{fig:close-approach}{Stellar radii of evolved stars, and closest
approaches between stars in destabilized systems.} Dashed red lines indicate the distribution
of closest approaches between the evolved star and another member
of an evolving triple  during the mass loss phase
(taken to be $\tau_{{\rm loss}}=0.5$ Myr). Solid blue lines show
the distribution of the stellar radii of the evolved mass losing star
at the point of instability. Inset shows the cumulative distributions.
Results were obtained from the three-body simulations of systems that
became unstable.}

\end{figure}

We define collisions in our three body integrations  as those cases where the
closest approach between the primary and another star is smaller than
the typical stellar radius of the evolving star, $R_{{\rm des}}$
where $R_{{\rm des}}$ is defined to be the stellar radius at the
point of instability, when the system first satisfies $Q<Q_{{\rm st}}$.
We only classify close approaches as collisions if they occur within
the 0.5 Myr mass loss timescale when the star is likely bloated. Figure
\ref{fig:close-approach} shows the cumulative distribution of closest
approaches amongst the unstable triples that we simulated. The distribution
of radii of the evolved stars that go unstable ($R_{{\rm des}}$
defined before) is also shown. Most stars involved
in the collisions are highly evolved AGB stars, with radii in the
range of tens to hundreds of AU.

We find the collision rate between the evolving star and either of
the other stellar components to be $f_{{\rm col}}\simeq0.09\pm0.03$
amongst the unstable systems, where the error bars are the 1 $\sigma$
statistical uncertainty. Note that this fraction is smaller than that obtained  from the simplified semi-analytic approximation
discussed in the appendix. However, detailed analysis of the distribution
of closest approaches between stars during the triple evolution, suggests
that the simple linear approximation in Eq. (\ref{eq:collision-prob})
is inaccurate; we find that the closest approach statistics depend more complexly on the 
various orbital properties of the system. We therefore adopt the collision
fraction found in our simulations for our subsequent calculations.

The sample chosen for the three body integrations excluded close inner binaries (with SMAs smaller than 15 AU),
because we did not account for mass transfer. Physical collisions are even more likely for these systems due to the monotonic
dependence of the close approach probability on the inner binary separation
(seen both in our integrations and in Eq. \ref{eq:collision-prob}.)  Thus the neglected systems likely have a higher collision probability.
Because we adopt the same collision fraction for more compact triples,
these rates should be considered a lower bound. Compact systems comprise
about half of our observed sample, so the neglect of these systems
in our calculations would change our results by at most a factor of
two.

Note that we cannot correctly follow
the full evolution of systems in which tidal effects dominate the
dynamics at any point. Nevertheless, for the short timescale of the
TEDI, tidal effects become important when the closest approach is
comparable to the stellar radius, and can be considered a sub-type
of collision. Moreover, because most collisions occur when the physical
radius is comparable to the tidal disruption radius, our neglect of
tidal effects should not significantly alter our calculated collision
rates. Determining the configurations of post-collision systems is beyond the scope of this work.

\section{Galactic Collision rate due to the TEDI}

\label{sec:rate} In the previous sections we provided estimates
for the fraction of destabilized systems, $f_{\rm des}$, and the fraction
of colliding systems, $f_{\rm col},$ among these systems. We can now
make use of these results to estimate the overall rate of stellar
collisions due to the TEDI. Taking the estimate of the collision
rate together with the calculated fraction of destabilized triples,
we find that the collision rate between evolved and companion stars/compact
objects in a Milky Way like galaxy, over the age of the Universe,
$t_{{\rm H}}$ is: \begin{eqnarray}
 & \Gamma_{{\rm col}} & =\frac{N_{\star}\times f_{{\rm evolve}}\times f_{{\rm triple}}\times f_{{\rm des}}\times f_{{\rm col}}}{t_{{\rm H}}}=\\
 & = & 1.2\times10^{-4}\left(\frac{N_{{*}}}{1.5\times10^{11}}\right)\nonumber \\
 & \times & \left(\frac{f_{{\rm evolve}}}{0.1}\right)\left(\frac{f_{{\rm triple}}}{0.1}\right)\left(\frac{f_{{\rm des}}}{0.012}\right)\left(\frac{f_{{\rm col}}}{0.09}\right)\,{\rm {yr}^{-1},\label{eq:collision-rate}}\nonumber \end{eqnarray}
where $N_{\star}$ is the number of stars in the
Galaxy (\citealp{bin+08}; taking the mean stellar mass to be $0.3$
M$_{\odot}$; \citealp{kro+01}), $f_{{\rm evolve}}$ is the observed
fraction of evolved stars (the fraction of WDs, \citealp{bel+02}
can serve as a proxy for their total number), $f_{{\rm triple}}$
is the fraction of triple systems (where we conservatively adopt the
triple fraction of $f_{{\rm triple}}=0.1$, found for F/G/K stars,
\citealp{rag+10}, for stars of all masses). We adopt $f_{{\rm des}}$
(the fraction of triple systems that become unstable) and $f_{{\rm col}}$
(the fraction of unstable systems in which a collision occurs) from
our calculations in the previous sections. 

For comparison, the collision rate in the GC M15, one of the densest
in the galaxy ($n_{\rm c}>10^{6}$ M$_{\odot}$ pc$^{-3}$; \citealp{umb+08}),
was estimated to be $\Gamma=2\pm1\times10^{-7}$ yr$^{-1}$ (\citealp{umb+08};
see also refs. \citealp{sha99,lee+10}). The total collision rate
in Galactic GCs, which is dominated by such clusters (i.e. over 100 GCs exist in the Galaxy, but the dominant contribution to collsions come from the most dense, massive clusters), is therefore
$\Gamma_{{\rm gal}}=4\times10^{-6}\,(N_{{\rm GC}}/20)$ yr$^{-1}$,
where ${\rm N}_{{\rm {\rm GC}}}\approx20$ is an estimate of the number
of similar GCs in the galaxy as massive and dense as M15 \citep{gne+97}.
Collisions due to destabilized triples are therefore about $30\,(N_{{\rm GC}}/20)$
times more frequent than collisions due to random encounters in GCs.
The number of extragalactic GCs is typically proportional to the host
galaxy mass \citep{bro+06}, suggesting a similar ratio of field to
GC collisions in other galaxies.

Note that the collsion rate due to TEDI \emph{inside} GCs scales with the total number of stars in GCs and does not depend on their specific density. Taking approximately 150 GCs exiting in the Galaxy with about $10^6$ stars in each cluster, we find the total collsion rate in GCs due to TEDI to be at most 0.1 percent of the total Galactic rate, and only about 3 percente of the collsion rate in GC due to random encounters. In other words, inside GCs, sellar collsions are dominated by random encounters rather than by triple stellar evolution.

\section{Consequences of the TEDI}

\label{sec:outcomes} 

The TEDI scenario has a wide variety of possible outcomes.
Many of these are very similar to those resulting from binary-single
and single-single close encounters between stars and/or compact objects
that occur in globular clusters. However, the TEDI scenario provides a pathway to these
exotic scenarios previously reserved for dense stellar populations. 

In the following we discuss various outcomes of the TEDI and
highlight their unique aspects, either in terms of their qualitative
difference from typical encounters in clusters, or their specific relations
to systems in the field. We first briefly overview several
potential outcomes, and then discuss in more detail a specific example,
providing a novel formation scenario for Sirius like eccentric WD
binaries.

\textbf{\emph{\underbar{Stellar collisions in the field:}}}\
As discussed above, the TEDI produces a high rate of stellar collisions
\emph{in the field}, most of which involve AGB
stars; these were previously thought to  comprise a
negligible fraction of all collisions. These encounters might be observed
as intermediate luminosity optical transients (with total energy ranging
between $10^{46}-10^{48}$ ergs; these could be detectable in current
and future optical transient surveys; \citealp{kul+09}), possibly
similar to those suggested to occur due to mergers or through tidal/off
axis collisions in eccentric binaries \citep{kas+10,smi11}. Beside
potentially producing immediate energetic transient events, the TEDI
can produce a wide variety of collision products, including merged
stars and blue stragglers as well as peculiar binaries.

Numerical simulations of close encounters with stars bloated to $100 \, R_{\odot}$
 show that the resulting collisional and/or
tidal captures lead to the formation of highly eccentric binaries
\citep{bai+99,yam+08}. Such collisions might also lead to a common
envelope phase and the formation of a close binary. The TEDI scenario
therefore predicts the existence of a sub-population of close, but
eccentric, evolved binaries in the field. The TEDI can explain the
puzzling orbits of WD binaries such as the nearby Sirius system (see
below), as well as highly eccentric Barium binary systems (expected
to form through mass transfer in a close binary). These systems pose
a problem for standard binary stellar evolution models, which produce
only \emph{circular}  close binaries \citep{bon+08,kar+00,izz+10} (but see \citeauthor{der+12} (2012), and references therein, for some alternative suggestions).
The peculiar close WD-blue
straggler binary (KOI-74) found by Kepler, which may have formed in a highly
eccentric evolved binary system \citep{dis11},  could easily be produced by the TEDI. Similarly, TEDI
evolution can lead to the formation of eccentric WD-WD binaries, which
could provide unique sources for future gravitational wave detection
missions, and were predicted to exist only in GCs \citep{wil+07}.
Our findings suggest that there could be a few times more such sources
in the field than estimated to exist in GCs.

\textbf{\emph{\underbar{Neutron stars/black hole binaries: }}} Mass
loss from primaries more massive than $8$ M$_{\odot}$ which undergo the TEDI, might form a close
eccentric binary before the supernova explosion of the evolving massive
star. Since such binaries have a higher probability of surviving the
SN explosion \citep{hil83}, the TEDI might aid in the retention of
neutron stars and black holes in binaries, and the later formation
of X-ray and pulsar binaries. Such triples could later evolve,
during mass loss/mass transfer from the other triple components. Possible
outcomes include instability and thus collisions with a neutron star/black
hole, or even exchanges. See \citet{fre+11} and \citet{por+11} for
the latter possibility, but note that these studies did not account
for the possibility of a physical collision during the chaotic evolutionary
phase.

\textbf{\emph{\underbar{Envelope stripping through collisions:}}} A
collision between a neutron star and its evolving companion could
result in the tidal disruption of the latter. The remnants of this
star might form an accretion disk surrounding the neutron star, which
could provide the necessary mass and angular momentum to create a
recycled pulsar \citep{ver+87}. If the third companion is ejected
during the course of the instability, the pulsar would appear as an
isolated recycled pulsar in the field. If the third companion is retained
it will be observed as a wide and likely eccentric recycled binary
pulsar in the field. The latter scenario provides an additional channel
for the formation of the pulsar J1903+0327 \citep{cha+08,fre+11,por+11}.
Another possibility is the formation of a recycled pulsar with a stripped
companion on an eccentric orbit, if the companion's core stays intact.

There are many observed peculiar stars and compact remnants
thought to form through envelope stripping by a close companion, such
as hot sub-dwarf stars \citep{han+02}, helium white dwarfs \citep{mar+95},
type Ib/c supernova progenitors \citep{yoo+10}. However, cases are
found where these peculiar objects are observed to be single or to
have a wide companion, configurations that are inconsistent with the
close binary progenitor scenario. Such cases might be explained
through a TEDI induced collision.
Further studies are required in order to quantitatively explore each scenario.

\section{Formation of the Sirius WD-binary system}

One of the closest WD binaries to Earth is the Sirius system composed
of a MS star of mass $2\, M_{\odot}$ and a WD companion of about
a Solar mass \citep{lie+05}. The SMA of 20 AU and and high eccentricity
of $0.59$ make this binary quite peculiar, as standard stellar evolution
scenarios suggest that the progenitor of such a close binary would
have circularized during stellar evolution of the system \citep{bon+08,izz+10}. 
However, the Sirius system could have formed through the TEDI; alternative scenarios are also in development (\citeauthor{der+12} 2012; P. Eggleton, private communication 2012). 

To assess the likelihood that Sirius formed via this mechanism, we
ran 6500 three-body simulations of the dynamical evolution of initially
stable triple systems undergoing mass loss. For these systems,  we follow the exact
mass loss and radius evolution as obtained from the SSE code. The
primary starts out with $M=5.05\, M_{\odot}$, and ends as a $0.98\, M_{\odot}$
WD. We evolve the systems analytically until the primary
has reached about $4$ M$_{\odot}$ and begins the rapid mass loss
phase. We assume that the secondary and tertiary components have undergone
no evolution. We explore a range of inner binary separations from
$15-30$ AU, outer binary separations from $3.5-5.5$ times the inner
separation, eccentricities from $0-0.25$, and companion masses from
$0.6-5.5\, M_{\odot}$. One component is always constrained to be
$2\, M_{\odot}$, the mass of Sirius A. The minimum separation of
the inner binary is chosen to avoid mass transfer. To account for
variations in the chaotic evolution at late times, we run each set
of parameters starting from five different initial orbital phases.
We limit the initial inclination of the systems to $i=0.1$ radian.

\begin{figure}
\centering{}\includegraphics[scale=0.45]{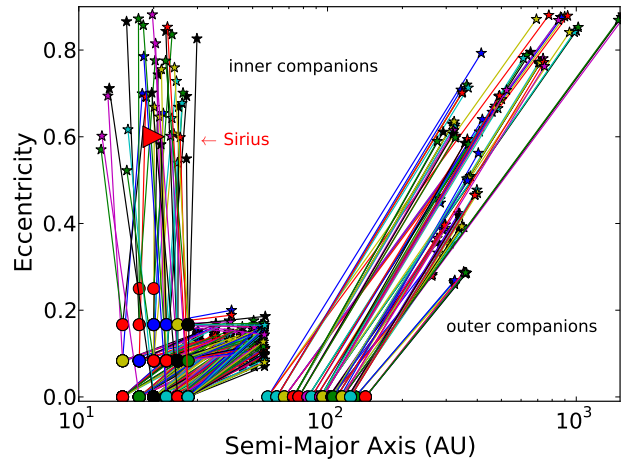} \caption{\label{fig:Sirius} The initial and final eccentricities and semi-major
axes of triple systems evolved using our three-body simulations. Circles
represent the initial orbital configuration of a given pair of stars
in the triple. Note that each circle corresponds not only to multiple
realizations of a given system, but also to different masses for the
third component. Lines connect the initial and final configurations,
which are marked by stars. Each initial configuration can correspond
to multiple final outcomes due to the chaotic nature of the instability.
Final configurations similar to the observed Sirius system (marked
with a red triangle), or even more eccentric are a typical outcome
for systems with initial SMAs in the range 15-30 AU. Also shown are
binaries formed via an exchange between the inner and outer companions,
followed by ejection of the initial inner companion.}

\end{figure}

In Fig. \ref{fig:Sirius}, we show examples of final binary configurations
of unstable systems where one of the stars was ejected and a Sirius
like binary was left behind. Also shown is the observed SMA and eccentricity
of the Sirius system. The Sirius binary falls within one of three
populated regions of final configurations, suggesting that Sirius
could have formed through the TEDI process. The inner cluster of eccentric
systems is formed via ejection of the wider tertiary companion, while
the outer cluster of eccentric systems formed via an exchange followed
by the ejection of the initial inner binary companion to the WD. The
moderately eccentric systems are composed of bound triple systems
that underwent only mild eccentricity excitation.

We note that if Sirius and other objects formed through this mechanism, it might
be possible to identify its ejected third star, as a
very wide companion (e.g. \citealp{jia+10}) or through similar methods
used to locate stars that originally formed close to the Sun \citep{por+09,bob+11}.

\section{Triple stellar evolution coupled with secular Kozai cycles}

\label{sec:koz} Triple systems in which the inner and outer orbits
have a high mutual inclination may secularly evolve through Kozai
cycles, in which the inclination and eccentricity of the inner binary
can periodically/quasi-periodically change by a large amplitude \citep{koz62,lid62},
and even flip from prograde to retrograde orbits (\citealp{nao+10};
and vice versa). It was already suggested that the coupling of secular
Kozai evolution with dissipative processes such as tidal friction
or GW emission could play an important role in the evolution of the
inner binaries (e.g. \citealp{maz+79,egg+06b,per+09,tho11} and references
therein). Mass loss evolution in triple systems could also couple
with this secular Kozai process, and change its outcomes. Our simulations
suggest that mass loss does not quench Kozai cyles. The high eccentricities
that could be induced through the Kozai mechanism
can therefore lead to close encounters between the components of the
inner binary during peri-center approach prior to destabilization
during the AGB stage. Kozai evolution could lead to close interactions
between triple components at a late epoch in the evolution, even if
the original triple did not interact while on the main sequence (see also \citeauthor{sha+12} 2012).
The coupled stellar-dynamical evolution TEDI stable, Kozai susceptible,
systems is beyond the scope of this paper, and will be explored elsewhere.

\section{Discussion and summary}

\label{sec:summary} In this paper we have studied the triple evolution
dynamical instability (TEDI), in which mass loss in an evolving triple
system leads to orbital instability. The chaotic evolution that ensues
gives rise to close encounters between the stars and/or exchanges
between them. Such evolution typically ends in a collision between
two of the stars or the ejection of one of the components.
This instability can lead to the dynamical formation of eccentric compact
binaries, and produces a high stellar collision rate between stars
in the field. The TEDI scenario results in a collision
rate in the field of approximately $10^{-4}$ yr$^{-1}$ per Milky-Way
Galaxy, about 30 times higher than the collision rate in all Galactic
GCs. This high rate is attributable to two features of the TEDI. First, the chaotic orbital evolution
 results in binary-single star like resonant encounters in the field,
which provide numerous opportunities for collisions. Secondly, the
cross section for physical collisions is much larger than in typical
random encounters. The mass loss that precipitates the instability
reaches its maximum at the AGB phase, and therefore the instability
typically occurs when the stellar radius is at its peak, a few hundred
Solar radii (as seen in Fig. \ref{fig:close-approach}). The cross
section for collisions is therefore increased proportionally. In contrast,
random collisions in GCs are heavily dominated by MS and slightly
evolved stars \citep{lom+06,umb+08} with radii of $1-10$ Solar radii.
The discrepancy between the peak collision phases is due to the $10^{5}-10^{6}$
year duration of the AGB phase. This timescale is a tiny fraction
of a star's life, and thus short compared to the random encounter
timescale, even in the densest environment. In contrast, the close
encounters during the TEDI always coincide with the AGB phase. The
TEDI model therefore requires not only a revision of the rate of stellar
collisions, but also a reconsideration of their nature; most stellar
collisions involve AGB stars. 

In addition to enhancing the number of stellar collisions,
the TEDI suggests that various types of exotic binaries collisionally formed inside GCs (X-ray
binaries, cataclysmic variables, single and binary recycled pulsars,
eccentric WD binaries, including eccentric inspiraling WD-WD GW sources,
and stripped stars; \citealt{fab+75,poo+03,bai+99,lom+06,wil+07} ) can similarly \emph{collisionally} form through TEDI \emph{outside} of dense clusters; adding an additional evolutionary channel for the formation of such binaries in the field.
In particular, the TEDI provides a novel explanation for the origin of
WD binaries observed in puzzling configurations, such as the well known
Sirius system. The exotic stellar systems produced through
these pathways have an outsize impact on our understanding of fundamental
physics, and the evolution of stars in the field. They serve as the
main sources for gravitational wave emission, and as unique probes
of the structure of compact objects.

This study focused on the TEDI scenario for triple
stellar systems. However, a system comprised of a stellar binary and
a planet on a circumstellar orbit around an evolving star can also
suffer a similar fate \citep{per10,per11}. The evolution of such
systems has recently been described by \citet{kra+12}. In addition,
the TEDI scenario can be extended to higher multiplicity systems (note
that $1/4$ of all multiple systems are quadruples or higher order
multiples \citealp{rag+10}).
Such systems can produce an even wider variety of outcomes than triple
systems, and in general mimic almost any type of outcome that is expected
to occur following close encounters in GCs.

\acknowledgments We thank J. Hurley for providing the openly accessible
SSE and BSE stellar evolution codes, S. Tremaine for helpful discussion
on wide binaries in the context of this work, as well as F. Rasio and P. Eggelton for helpful
comments on an earlier version of this manuscript. We also thank the referee, Christopher Tout, for helpful
suggestions which much improved the clarity of this work. HBP acknowledges
support from the BIKURA (FIRST) Israel Science Foundation and the
CfA fellowship through the Harvard-Smithsonian Center for Astrophysics.
KMK is supported by the Institute for Theory and Computation Fellowship
through the Harvard College Observatory.

\appendix{}

\section{Simplified estimate of the rate of stellar collisions in the TEDI
scenario}

The frequency of close approaches for initially unstable systems has
been characterized in previous studies (see \citealp{sas+74,val+06}
and references therein; see also \citealp{sig+93} for the related
issue of binary-single encounters). Triples starting
from unstable configurations 
undergo similar evolution to that observed in simulations of low velocity,
resonant encounters between a binary and a single star \citep{val+06}. The evolution of unstable triples can thus be modeled as a set of
random encounters between an incoming third body and an inner binary.
After any encounter, the incoming star can be ejected from the system,
go through an exchange with one of the binary components, or collide
with one of them. If a collision or ejection does not occur on a given
passage, the cycle repeats, and the system experiences another encounter
(though the components of the inner binary may change in the event
of an exchange). 
These encounters cease when a collision or ejection occurs (a long-lasting
stable triple can only be formed via energy dissipation, e.g. following
a collision or a strong tidal encounter,).  Previous three body integrations
show that, on average, approximately $60$ such encounters occur before
a star is ejected \citep[see][and references therein]{val+06}, thus
the probability for a close approach between any two stars in the
system is proportionally increased compared with single-single stellar
encounters. 
 These results provide a simple
estimate for the collision rate. During the chaotic evolution of unstable
triple systems, the probability, ${\rm P}(<r_{\rm close},\, a_{{\rm out}})$
for a close approach of $r_{\rm close}\ll a_{{\rm out}}$ or smaller is
\citep{val+06}: \begin{eqnarray}
 & {\rm P}(<r_{close},\, a_{{\rm out}})\sim240\left[1+1.875\left(\frac{L/L_{{\rm max}}}{0.5}\right)^{2}\right]\times\frac{r_{\rm close}}{a_{{\rm out}}}\nonumber \\
 & \simeq0.11\left[1+1.875\left(\frac{L/L_{{\rm max}}}{0.5}\right)^{2}\right]\times\left(\frac{r_{\rm close}}{100{\rm R}_{\odot}}\right)\left(\frac{a_{{\rm out}}}{1000\,{\rm AU}}\right)^{-1},\label{eq:collision-prob}\end{eqnarray}
 where $0\le{\rm L/L}_{{\rm max}}\le1$ is the normalized angular
momentum of the system ($L_{max}$ is the angular momentum for a circular
orbit of the same energy). For the typical system that becomes unstable,
 Eq. (\ref{eq:collision-prob}) gives a probability of order
unity because the stars are often 100's of solar radii at the onset
of instability; however, the time spent at this radius, $t_{{\rm R,des}}$,
may be shorter than the typical length of the chaotic evolution, $t_{{\rm chaotic}}\approx60\times p_{{\rm out}}^{{\rm des}}$,
where $p_{{\rm out}}^{{\rm des}}$ is the period of the triple outer
orbit at the point of destabilization. Thus the probability for a
collision is quenched by a factor of approximately $\min(t_{{\rm R,des}}/60p_{{\rm out}}^{{\rm des}},1)$.

Applying this correction factor, we can calculate the collision fraction
for our observed sample. We take $R_{\rm des}$ to be the stellar
radius of the AGB star at the point of destabilization, $a_{{\rm out}}$ to be the outer binary separation,
and $t_{{\rm R,des}}$ to be the AGB lifetime at this radius or above. All these parameters
are calculated separately for each system in our observed triples
sample, at the point of instability, using the models described above
(Section \ref{sec:unstable}).

According to this calculation, the fraction of destabilized triples that suffer a direct
collision between the evolving star and one of the other triple components
is $0.49\pm0.03$ for the observed sample of triples (with the assumed eccentricity distribution). Comparison with our numerical results suggests that the analytic calculation significantly overestimates collisions for TEDI systems.

\bibliographystyle{apj}

\end{document}